\documentclass[]{interact}

\usepackage{epstopdf}

\usepackage[numbers,sort&compress]{natbib}
\bibpunct[, ]{[}{]}{,}{n}{,}{,}
\makeatletter
\def\NAT@def@citea{\def\@citea{\NAT@separator}}
\makeatother

\begin{document}

\articletype{Special Issue "Nano-polycrystalline diamond and its applications"}

\title{Recent progress in high-pressure X-ray absorption spectroscopy studies at the ODE beamline }

\author{
\name{	Lucie Nataf\textsuperscript{a},
	Fran\c{c}ois Baudelet\textsuperscript{a}\thanks{CONTACT Fran\c{c}ois Baudelet. Email: francois.baudelet@synchrotron-soleil.fr Address: Synchrotron SOLEIL, l'Orme des Merisiers, Saint-Aubin, BP 48, 91192 Gif-sur-Yvette, France},
    Alain Polian\textsuperscript{a}, 
	Inga Jonane\textsuperscript{b},
  Andris Anspoks\textsuperscript{b},
  Alexei Kuzmin\textsuperscript{b}\thanks{Alexei Kuzmin. Email: a.kuzmin@cfi.lu.lv Address: Institute of Solid State Physics, University of Latvia, Kengaraga street 8, LV-1063 Riga, Latvia} and Tetsuo Irifune\textsuperscript{c}}
\affil{\textsuperscript{a}Synchrotron SOLEIL, l'Orme des Merisiers, Saint-Aubin, Gif-sur-Yvette, France; \textsuperscript{b}Institute of Solid State Physics, University of Latvia, Riga, Latvia;
\textsuperscript{c}Geodynamics Research Center, Ehime University, Matsuyama, Ehime, Japan }
}

\maketitle

\begin{abstract}
High-pressure energy-dispersive X-ray absorption spectroscopy is a valuable structural technique, especially, when combined with a nano-polycrystalline diamond anvil cell. 
Here we present recent results obtained using the dispersive setup of 
the ODE beamline at SOLEIL synchrotron.  The effect of pressure and 
temperature on the X-ray induced photoreduction is discussed on 
the example of nanocrystalline CuO. 
The possibility to follow local environment changes during pressure-induced phase transitions is demonstrated for $\alpha$-MoO$_3$ 
based on the reverse Monte Carlo simulations.
\end{abstract}


\section{Introduction}

An alternative to the traditional pattern of beamlines dedicated to the X-ray absorption spectroscopy, based on a two-crystal scanning mode, is given by the use of dispersive optics \cite{Fontaine1992,Blank1992,DAcapito1992,Dent1992}. In this technique, a bent crystal is used as a monochromator, and the setup is usually called Energy Dispersive EXAFS (EDE).
The continuous change of the Bragg incidence along the bent crystal opens an energy range in the reflected beam. The correlation between position and energy of the X-ray is exploited thanks to a position sensitive detector. The setup can be used to acquire both the  X-ray absorption near edge structure (XANES) as well
as extended X-ray absorption fine structure (EXAFS).

The EDE beamline at SOLEIL synchrotron, called ODE for Optic Dispersive EXAFS  \cite{ODE2011}, has extensively developed high-pressure XAS (X-ray Absorption Spectroscopy) and XMCD (X-ray Magnetic Circular Dichroism) techniques. The XMCD activity was already presented in \cite{Baudelet2016xmcd}.  Here we report on some high-pressure measurements done with a nano-polycrystalline diamond anvil cell (NDAC) \cite{Irifune2003}.

The main advantages of dispersive XAS are the focusing optics, the short acquisition time, and the great stability during the measurements due to the absence of any mechanical movement. This advantage allows the study of small samples, mandatory in the case of high-pressure studies, where the sample chamber is a hole in a gasket sandwiched between two diamonds. The EDE technique is particularly adapted for deglitching due to its live measuring process with classical diamonds, however, there are some limits in the 10 keV range. The use of NDAC allows one to  improve significantly the quality of the experimental signal eliminating spurious contributions from the diamond Bragg reflections \cite{Ishimatsu2012}. This opens the route of many new possibilities as seen in the following contributions of XAS pressure measurements.

\section{X-ray absorption spectroscopy at high-pressure}

\subsection{XANES as a probe of X-ray photoreduction}

High-intensity X-ray beams at modern sources of synchrotron radiation  
can cause radiation damage to a material. The effect is well known in 
the case of metal-organic complexes \cite{Olieric2007,Holton2009,Kubin2018}. 
This problem may be of particular importance for beamlines with the energy-dispersive setup when a polychromatic X-ray beam with the spectral range of several hundred electron-volts is focused to the spot of several tens of microns on the sample located in the diamond anvil cell. In particular, when high-pressure measurements are performed, the indirect damage of the sample 
placed in a solution, which plays the role of the pressure-transmitting medium, can  occur due to the radiolysis of the latter. 
In the radiolysis process, reducing radicals such as solvated electrons and hydrogen atoms are produced upon X-ray irradiation of the solution and are responsible for the sample change \cite{Jonah1995,Caer2011}. For example, the radiolysis can lead to a reduction of metal ions in aqueous solutions and is used for synthesis of metal nanoparticles \cite{Joshi1998,Lee2003,Oyanagi2014}.

Recently, we have demonstrated X-ray induced photoreduction of nanocrystalline CuO, and its dependence on crystallite size, temperature and pressure \cite{Kuzmin2018cuo}. Nanocrystalline CuO (nano-CuO) powder samples with the average crystallite size of 8~nm and 20~nm were prepared by a decomposition of Cu(OH)$_2$ precipitate in air at 130~$^\circ$C and 150~$^\circ$C, respectively.  The Cu K-edge XANES spectra were collected in the pressure range of 0-23~GPa and the temperature range of 
10-300~K. The sample pressure  and temperature were controlled using  a membrane-type nano-polycrystaline diamond anvil cell (NDAC) \cite{Irifune2003,Ishimatsu2012} and  liquid helium cryostat. The use of NDAC allowed us to accumulate the  experimental data free from Bragg peaks due to the diamonds. The polychromatic photon flux on the sample was about 10$^{9}$ photons/s/eV in 25$\times$35~$\mu$m FWHM.

\begin{figure}
\centering
\includegraphics[width=0.95\linewidth]{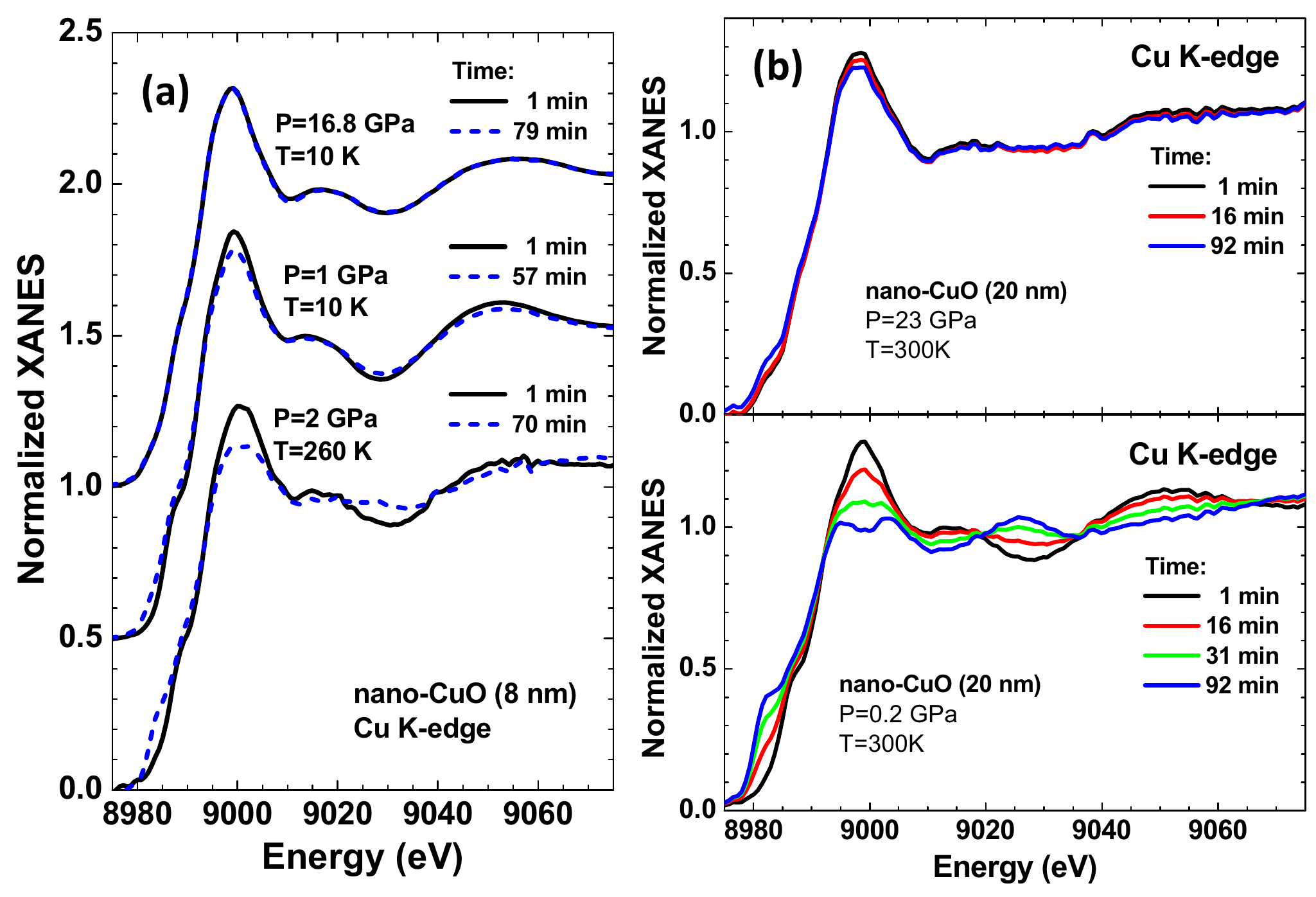}
\caption{(a) Temperature, pressure and time dependence of the Cu K-edge XANES of nano-CuO (8~nm). Solid line -- starting spectrum, dashed line -- spectrum after long time (57-79~min) exposition. Spectra are shifted vertically for clarity.
(b) Pressure and time dependence of the Cu K-edge XANES of nano-CuO (20~nm). Upper panel -- P=23~GPa, lower panel -- P=0.2~GPa.} 
\label{fig1}
\end{figure}

The sensitivity of the Cu K-edge XANES in nano-CuO (8~nm) to temperature and pressure is shown in Fig.~\ref{fig1}(a).  
The emergence of metallic copper is visible at $T$=260~K and $P$=2~GPa 
after 70~minutes of exposure to X-rays as an increasing shoulder at 8983~eV and a decreasing main peak at 9000~eV. X-ray induced photoreduction 
occurs more rapidly at higher temperatures and lower pressure.
For larger crystallite size (20~nm), the reduction process takes longer time (Fig.~\ref{fig1}(b)). At $T$=300~K and $P$=0.2~GPa, copper oxide is fully converted into metallic copper after about 90~minutes of irradiation. However,
an increase of pressure to 23~GPa stabilizes the oxide phase. 
Note that the photoreduction was not observed for microcrystaline CuO 
at similar conditions. 
Thus, the rate of nano-CuO photoreduction to metallic copper increases with decreasing nanoparticle size but can be reduced by a decrease of temperature or an increase of pressure. 
Performing experiment at low temperatures decreases 
the mobility of reducing radicals \cite{Corbett2007,Meents2010}, whereas 
an increase of pressure results in CuO nanoparticles agglomeration
thus restricting their free surface and impeding reduction. 
The obtained results suggest that 
possible radiation damage should be taken into account in experiments with high-flux X-ray beams, especially, in the case of nanosized materials.

\subsection{Pressure-induced phase transitions probed by EXAFS}

The sensitivity of EXAFS to the local atomic structure of a material and recent developments 
in the EXAFS data analysis based on the reverse Monte Carlo (RMC) simulations \cite{Timoshenko2014rmc} provide an invaluable tool to follow 
pressure-induced phase transitions. However, high-quality experimental 
data remain the main limiting factor to unleash the potential of the method.

\begin{figure}
	\centering
	\includegraphics[width=0.95\linewidth]{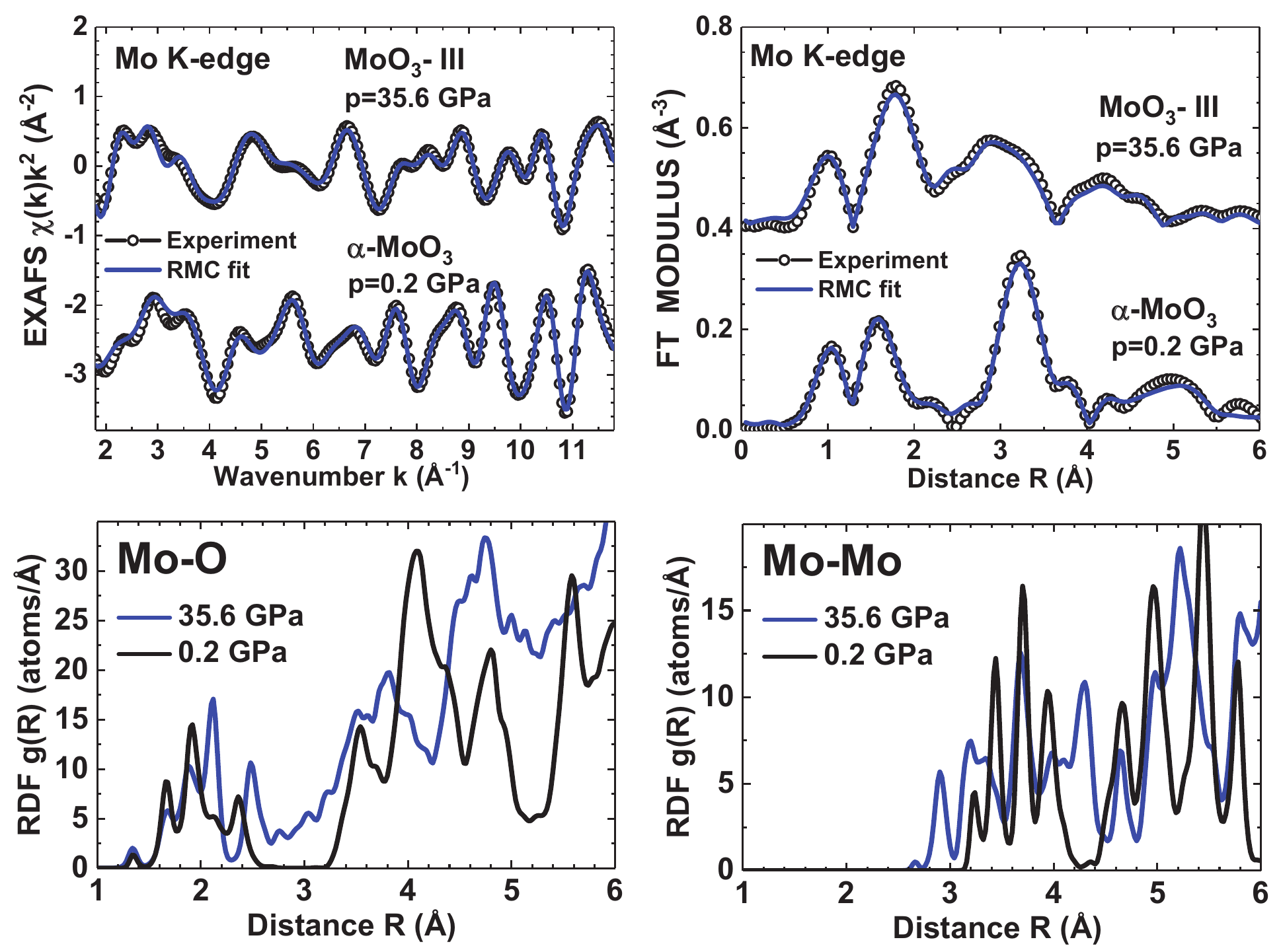}
	\caption{Results of the RMC-EXAFS calculations for $\alpha$-MoO$_3$ (0.2~GPa) and MoO$_3$-III (35.6~GPa) phases. Upper row: comparison of the experimental and calculated Mo K-edge EXAFS spectra $\chi(k)k^2$ and their Fourier transforms. Curves are shifted vertically for clarity. Lower row: the radial distribution functions (RDFs) for Mo--O and Mo--Mo atom pairs reconstructed by the RMC method. } 
	\label{fig2}
\end{figure}

The pressure-induced (up to 36~GPa) transformations  in 2D layered oxide  $\alpha$-MoO$_3$ were studied at the Mo K-edge in \cite{Kuzmin2019moo3}. Good quality experimental EXAFS data were acquired using the NDAC cell and allowed us to perform analysis up to 6~\AA\  using the RMC method. 
The structural models obtained by RMC give the Mo K-edge EXAFS spectra 
in good agreement with the experimental ones (Fig.~\ref{fig2}). The corresponding atomic 
coordinates were used to calculate the radial distribution functions $g(R)$ for Mo--O and Mo--Mo atom pairs as a function of pressure and allowed us to follow the details of local structure transformations upon phase transitions (Fig.~\ref{fig2}).  

At room temperature,  molybdenum trioxide  exhibits  two phase transitions upon compression in the range of 0-43~GPa \cite{Liu2009}.  $\alpha$-MoO$_3$ transforms to monoclinic MoO$_3$-II phase ($P2_1/m$) at $\sim$12~GPa,  and, next, to monoclinic MoO$_3$-III phase ($P2_1/c$) at $\sim$25~GPa. The first two phases  ($\alpha$-MoO$_3$ and MoO$_3$-II) have layered structure composed of strongly distorted  MoO$_6$ octahedra, whereas a collapse of the interlayer gap occurs in MoO$_3$-III phase \cite{Liu2009}. 

The difference between $\alpha$-MoO$_3$ and MoO$_3$-III phases is well established by EXAFS. 
The transition is accompanied by a change of the Mo--O--Mo angles between neighbouring molybdenum-oxygen octahedra from 
$\sim$142$^\circ$ and $\sim$168$^\circ$ in $\alpha$-MoO$_3$ \cite{Asbrink1988} to $\sim$149$^\circ$ in MoO$_3$-III \cite{Liu2009}. The disappearance of the Mo--O--Mo angle equal to 168$^\circ$ is responsible for a decrease of the second shell peak (at $\sim$3.2~\AA\ in Fig.~\ref{fig2}) amplitude in the Fourier transform of the EXAFS spectrum
of MoO$_3$-III due to a reduction of the multiple-scattering (MS) effects within the Mo--O--Mo atomic chains.

The reconstruction of the local environment around molybdenum using the reverse Monte Carlo (RMC) method \cite{Timoshenko2014rmc} shed light on the 
pressure dependence of the radial distribution functions (RDFs) for Mo--O and Mo--Mo atom pairs in details.

At  small pressure of about 0.2~GPa, there are three groups of two oxygen atoms each located at $\sim$1.70, 1.96 and 2.26~\AA, which form strongly distorted MoO$_6$ octahedra  in $\alpha$-MoO$_3$.  At high pressure (35.6~GPa), the collapse of layered structure leads to an increase of molybdenum coordination. Six nearest oxygen atoms from the same layer are responsible for the peaks at $\sim$1.68, 1.88 and 2.12~\AA, whereas the 7th oxygen atom bridging two layers is located at $\sim$2.48~\AA. The high pressure influences also the Mo--Mo distribution, leading to a shortening of the distance between neighboring molybdenum-oxygen polyhedra connected by edges (peaks at 2.9 and 3.25~\AA). 
Thus, the change of the molybdenum local environment upon phase transitions is well evidenced by the RMC analysis of high-pressure Mo K-edge EXAFS spectra.

\section{Conclusion}

In this paper, we describe the potential of high-pressure energy-dispersive X-ray absorption spectroscopy (XANES/EXAFS) in combination with a nano-polycrystalline diamond anvil cell. 
The experimental setup is well suited for such measurements, making it possible 
to obtain  experimental data of good quality, being free 
from the Bragg contribution from diamonds. At the same time, 
the accessible spectral range of the dispersive setup is restricted by the linear size of the detector, which imposes limitations on the resolution in real space. In addition, high intensity of X-rays focused on a sample can cause radiation damage. 

Two examples of high-pressure XAS studies of transition metal oxide compounds (CuO and MoO$_3$) are discussed. They illustrate the sensitivity of copper oxide nanoparticles to reduction under X-ray irradiation and the ability  to track local structural changes during phase transition in a 2D layered-type molybdenum trioxide using an advanced analysis method based on the reverse Monte Carlo algorithm.

\section*{Disclosure statement}

No potential conflict of interest was reported by the authors.

\section*{Funding}

A.K. and I.J. are grateful to the Latvian Council of Science project no. lzp-2018/2-0353 for financial support.   The  research  leading  to  these results has been partially supported by the project CALIPSOplus under the Grant Agreement No. 730872 from the EU Framework Programme for Research and Innovation HORIZON 2020.

\end{document}